\documentclass[useAMS,usenatbib,usegraphicx]{mn2e}
\usepackage{amsmath}
\usepackage{graphicx}

\title[Automated classification of periodic variable stars]{Improved methodology for the automated classification of periodic variable stars}
\author[J. Blomme et al.]
 {J.~Blomme,$^1$ L.M.~Sarro,$^2$ F.T.~O'Donovan,$^3$ J.~Debosscher,$^1$ T.~Brown,$^{4,5}$
\newauthor M.~Lopez,$^6$ P.~Dubath,$^7$ L.~Rimoldini,$^7$ D.~Charbonneau,$^8$ E.~Dunham,$^9$
\newauthor G.~Mandushev,$^9$ D.R.~Ciardi,$^{10}$ J.~De Ridder,$^1$ C.~Aerts$^{1,11}$ \\
$^1$ Instituut voor Sterrenkunde, Katholieke Universiteit Leuven, Celestijnenlaan 200D, B-3001 Leuven, Belgium \\
$^2$ Dpt. de Inteligencia Artificial, UNED, Juan del Rosal, 16, 28040 Madrid, Spain \\
$^3$ California Institute of Technology, 1200 E. California Blvd. Pasadena, CA 91125, USA \\
$^4$ Las Cumbres Observatory Global Telescope, Goleta, CA 93117, USA \\
$^5$ Departement of Physics, University of California, Santa Barbara, CA 93106, USA \\
$^6$ Centro de Astrobiolog\'{\i}a (INTA-CSIC), Departamento de Astrof\'{\i}sica, PO Box 78, E-28691, Villanueva de la Ca\~nada, Spain \\
$^7$ Geneva Observatory, Chemin d'\'Ecogia 16, CH-1290 Versoix, Switzerland \\
$^8$ Harvard-Smithsonian Center for Astrophysics, 60 Garden Street, MA 02138, USA \\
$^9$ Lowell Observatory, 1400 W. Mars Hill Road, Flagstaff AZ 86001, USA \\
$^{10}$ NASA Exoplanet Science Institute/Caltech, Pasadena, CA 91125, USA \\
$^{11}$ Department of Astrophysics, IMAPP, Radboud University Nijmegen, POBox 9010, 6500 GL Nijmegem, the Netherlands}

\begin{document}

\maketitle
\begin{abstract}
We present a novel automated methodology to detect and classify periodic variable stars in a large database of photometric time series.
The methods are based on multivariate Bayesian statistics and use a multi-stage approach. We applied our method to the ground-based data 
of the TrES Lyr1 field, which is also observed by the \textit{Kepler} satellite, covering $\sim 26\,000$ stars. We found many eclipsing
binaries as well as classical non-radial pulsators, such as slowly pulsating B stars, $\gamma$ Doradus, $\beta$ Cephei and $\delta$ Scuti 
stars. Also a few classical radial pulsators were found.
\end{abstract}


\section{Introduction}
In recent years there has been a rapid progress in astronomical instrumentation giving us an enormous amount of new time-resolved
photometric data, resulting in large databases. These databases contain many light curves of variable stars, both of known and unknown
nature. Well-known examples are the large databases resulting from the CoRoT \citep{Fridlund:2006} and Kepler \citep{Gilliland:2010}
space missions, containing respectively $\sim 100\,000$ and $\sim 150\,000$ light curves so far. The ESA Gaia mission, expected to be
launched in 2012, will monitor about one billion stars during five years. Besides the space missions, also large scale photometric 
monitoring of stars with ground-based automated telescopes deliver large numbers of light curves. The challenging task of a fast and
automated detection and classification of new variable stars is therefore a necessary first step in order to make them available for 
further research and to study their group properties.

Several efforts have already been made to detect and classify variable stars. In the framework of the CoRoT mission, a procedure for
fast light curve analysis and derivation of classification parameters was developed by \citet{Debosscher:2007}. That algorithm
searches for a fixed number of frequencies and overtones, giving the same set of parameters for each star. The variable stars were
then classified using a Gaussian classifier \citep{Debosscher:2007,Debosscher:2009} and a Bayesian network classifier \citep{Sarro:2009}.

In this paper we present a new version of this method to detect and classify periodic variable stars. In contrast to the previous
versions, the new automated methodology only uses significant frequencies and overtones to classify the variables with it giving less
rise to confusion, especially when dealing with ground-based data. In order to be able to deal with a variable number of parameters, we
also introduce a novel multi-stage approach. This new methohology offers much more flexibility. We applied this method to the
ground-based photometric data of the TrES Lyr1 field, covering about $\sim 26\,000$ stars. The classification algorithm considers
various classes of non-radial pulsators, such as $\beta$ Cep, slowly pulsating B (SPB) stars, $\delta$ Sct and $\gamma$ Dor stars, as
well as classical radial pulsators (Cepheids, RR Lyr) and eclipsing binaries (see, e.g., \citet{Aerts:2010} for a definition of the
classes of these pulsators).

\section{A new methodology}
\subsection{Variability detection}\label{sec:vd}
To detect and extract the variables we performed an automated frequency analysis on all time series. The algorithm first checks for a
possible polynomial trend up to order 2 and subtracts it, as it can have a large detrimental influence on the frequency spectrum
through aliasing. The order of the trend was determined using a classical likelihood-ratio test. Although the coefficients of the trend
are recomputed each time a new oscillation frequency is added to the fit, the order of the trend remains fixed.

After detrending, the algorithm searches for significant frequencies and overtones in the residuals, using Fourier analysis. 
The algorithm searches for the frequency with the highest amplitude in the discrete Fourier transform and checks if this period is 
significant, using the false alarm probability (\citet{HB:1986}, \citet{SC:1998}). Note that a detected frequency peak can be 
significant but unreliable. Reliability is checked through pre-specified frequency intervals that are not trustworthy (e.g. around
multiples of 1 c/d for ground-based data). Unreliable frequencies are prewhitened, but flagged as ``unreliable'' and are not used for
classification. If the frequency is the first significant reliable frequency, then the algorithm checks whether half of this frequency
is also significant and reliable. In this case, the original new frequency is replaced with half of this frequency to better model
the binary light curves. In a next step, the algorithm searches for significant overtones, using the likelihood-ratio test, to model 
possible non-sinusoidal variations (like those of RR Lyr stars). This procedure is repeated as long as significant frequencies are
found. These frequencies $\nu_n$ can be used to make a harmonic best fit to the light curve of the form:
\begin{equation}
\begin{split}
 f(t)= \sum_{i=0}^{K} a_i (t-t_0)^i\\ 
+ \sum_{n=0}^{N} \sum_{m=1}^{M} b_{n,m} \sin{(2\pi \nu_{n}m(t-t_0))}\\
+ c_{n,m} \cos{(2\pi \nu_{n} m(t-t_0))},
\end{split}
\end{equation}
\noindent with $0 \leqslant K \leqslant 2$ the order of the trend, and N, the number of significant frequencies, determined using the
false alarm probability and $M \geqslant 1$ the number of harmonics, determined using the likelihood-ratio test.

The frequency analysis method used by \citet{Debosscher:2007} performs well on properly-reduced satellite data for which it was designed, 
but not on noisier ground-based data, as many insignificant frequencies and overtones can degrade the performance of the classifier.

\subsection{The classifier}
The aim of supervised classification is to assign to each variable target a probability that it belongs to a particular predefined
variability class, given a set of observed parameters. This set of parameters (also called attributes) is obtained from the
variability detection pipeline described above and contains frequencies, amplitudes and phase differences. The classifier relies on a 
set of known examples, the so-called training set, of each class that needs to represent well the entire variability class.

We used a novel multi-stage approach, where the classification problem is divided into several sequential steps. This classifier
partitions the set of given variability classes ${C_i}$, into two or more parts: $\mathcal{C}^{(1)}$, $\mathcal{C}^{(2)}$, \dots. This
simplifies the classification by degrading the level of detail to a smaller number of categories. Each of these partitions 
$\mathcal{C}^{(i)}$, which can contain several variability classes, is then again splitted into $\mathcal{C}^{(i,1)}$, 
$\mathcal{C}^{(i,2)}$, \dots, which in turn can be partitioned into $\mathcal{C}^{(i,j,1)}$, $\mathcal{C}^{(i,j,2)}$, and so on, each 
time specializing the classification until each subpartition contains only one variability class. These partitions can be represented 
in a tree.

This approach offers several advantages compared to a single-stage classifier. The main advantage of this approach is that in each 
stage a different classifier and a different set of attributes can be used. This is important as attributes carrying useful
information for the separation of two classes can be useless or even harmful for distinguishing other classes. In each stage,
informationless attributes for the separation of the classes of interest can be removed, thereby significantly reducing possible
confusion. In addition, it is also possible to have a variable number of attributes. This allows to make different branches for mono-
versus multi-periodic pulsators. In this way we do not need a fixed set of attributes, thereby avoiding the introduction of spurious 
frequencies or overtones, which was sometimes the case in \citet{Debosscher:2007}. As already mentioned earlier, this too is important 
as insignificant attributes can degrade the performance of the classifier.

We took each of the classifier nodes in the multi-stage tree as a Gaussian mixture classifier. The Gaussian mixture classifier is
based on the general law of Bayes:
\begin{equation}
 P(C=c_i|A=a) = \frac{L(A=a|C=c_i)P(C=c_i)}{\displaystyle{\sum_{i=1}^{N_c}L(A=a|C=c_i)P(C=c_i)}},
\label{bayes}
\end{equation}
\noindent with $N_c$ the number of different classes. These classes can correspond to the variability classes (e.g. $\beta$ Cep,
SPB,...), but as the Gaussian mixture classifier is used at the nodes in the multi-stage classifier, a class in this context may 
also correspond to a group of variability classes relevant for a particular node. $P(C=c_i|A=a)$ is the a posteriori probability of 
the target belonging to class $c_i$ given the observational evidence $a$, and is the goal of the classification problem. 
$L(A=a|C=c_i)$ is the conditional likelihood of a attribute set $a$ given that it belongs to variability class $c_i$. $P(C=c_i)$ is
the a priori probability of a target belonging to class $c_i$. As no reliable prior values for variability classes are known yet,
we used a uniform prior.

\begin{figure*}
\includegraphics[width=16cm]{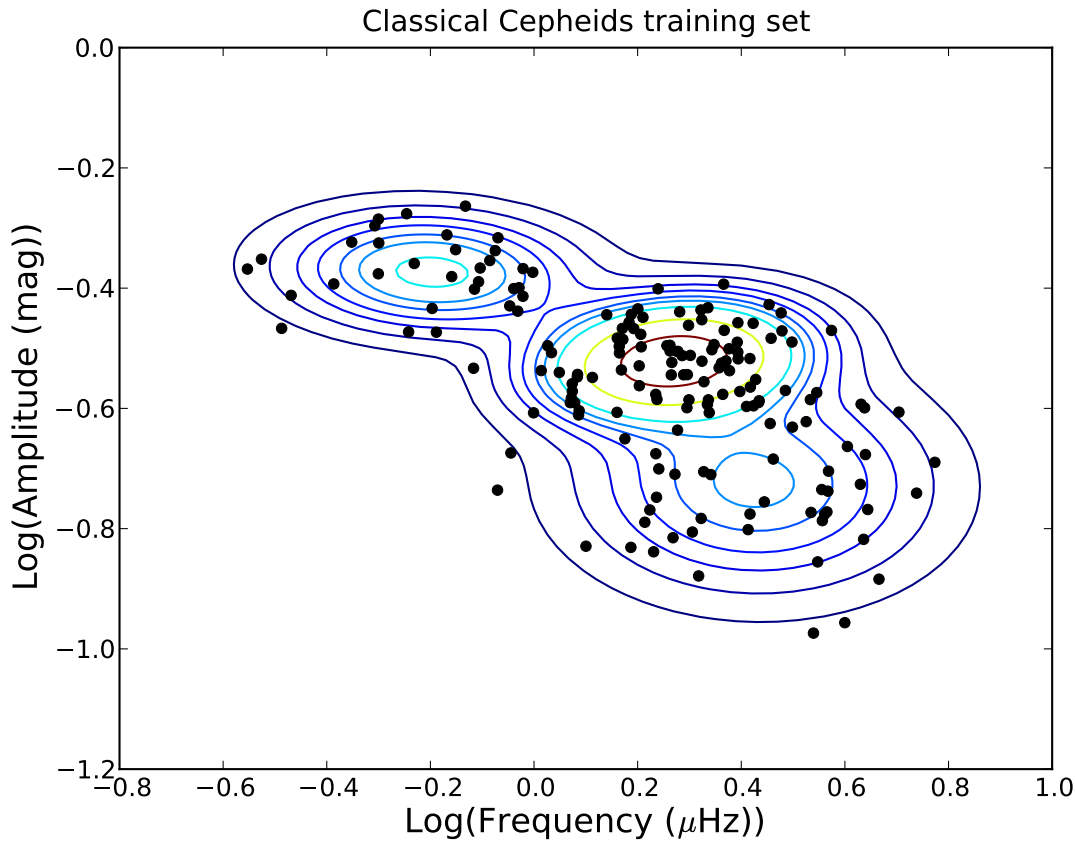}
 \caption{Gaussian mixture in the 2-parameter space ($log(\nu_1),log(a)$) for the Classical Cepheids in the training set estimated by
the Expectation-Maximization algorithm (EM).}
 \label{gm}
\end{figure*}

In previous versions of the classifier, the likelihood was approximated as a single Gaussian. Some of the variability classes, however,
are not well modeled by a single Gaussian. An example of this is shown in Fig. \ref{gm}, in which multiple components are clearly
preferable.

The likelihood is now approximated as a finite sum of multivariate Gaussians:
\begin{equation}
L(A=a|C=c_i)=\displaystyle{\sum_{k=1}^{M_i} \alpha_k \phi_k(a|\mu_k,\Sigma_k)},
\label{likelihood}
\end{equation}
\noindent where
\begin{equation}
\begin{split}
\phi_k(a|\mu_k,\Sigma_k)=\frac{1}{(2\pi)^{N_a/2}|\Sigma_k|^{1/2}}\\
exp\left(-\frac{1}{2}(a-\mu_k)'\Sigma_{k}^{-1}(a-\mu_k)\right),
\label{normal distribution}
\end{split}
\end{equation}
\noindent with $M_i$ the finite number of Gaussian components of class $c_i$, $N_a$ the number of attributes, $\alpha_k$ the a priori
probability to belong to component $k$, and $\mu_k$ and $\Sigma_k$, the mean vector and covariance matrix of Gaussian component $k$.

For each node of the multi-stage tree, the set of variability classes is partitioned. The best attributes are selected for that node
and the classifier is trained, meaning that the Gaussian mixture for each class is determined. To do so, we used the 
Expectation-Maximization (EM) method (see e.g. \citet{GM:1993}). Given a variability class, the unknowns are the number of Gaussian 
components of each class, the prior probability to belong to a particular component, and the mean vectors $\mu_k$ and covariance
matrices $\Sigma_k$ of each component. The EM algorithm is an iterative method for calculating maximum likelihood estimates of 
parameters in probabilistic models, where the model depends on unobserved latent variables. EM alternates between performing an
expectation (E) step, which computes the expectation of the log-likelihood evaluated by using the current estimate for the latent 
variables, and a maximization (M) step, which computes parameters maximizing the expected log-likelihood found in the E step. These
parameter-estimates are then used to determine the distribution of the latent variables in the next E step. Given the number of 
Gaussian components $N_c$, the remaining unknowns in the model can be determined by using this procedure. The actual number of
components is determined using the Bayesian information criterion (BIC), which is a criterion for model selection among a set of 
parametric models with different number of parameters. We obtained three components in the example in Fig. \ref{gm} using BIC.
The Akaike information criterium (AIC) gives the same number of components. This solution turns out to be very stable when changing
initial values, in the sense that the EM algorithm always converges to the same solution.

\subsection{Automated classification}
Once the classifiers in each node are trained, the targets can be classified. In each node we assign a probability to each target that
it belongs to a particular class relevant for that node. In order to obtain the final probability for each variability class we 
multiply the probabilities along the corresponding root-to-leaf path using the chain rule of conditional probability. Let 
$\mathcal{C}^{(k,j,\dots,l,n,m)}$ be the subpartition that contains only class $C_i$. The probability that the target $T$ belongs to 
$C_i$ is thus given by:
\begin{equation}
\begin{split}
 P(T \in C_i|\{A_i\})= \\
P(\mathcal{C}^{(k,j,\dots,n,m)}|\mathcal{C}^{(k,j,\dots,n)}) \dots P(\mathcal{C}^{(k,j)}|\mathcal{C}^{(k)}) P(\mathcal{C}^{(k)}),
\end{split}
\end{equation}
\noindent where we dropped ``$T \in$'' and the observed attributes $\{A_i\}$ in the right-hand side of the equation, for the sake of 
notational simplicity. We retain the most probable class assignment for a given variable star of unknown type and label it according
to the Mahalanobis distance.

Note that the denominator in Eq. \eqref{bayes} enforces the target to belong to one of the predefined classes, although the target can 
be very far from the class centers in attribute space. For that reason it is important to include an outlier detection step to flag
possible wrong predictions. \citet{Debosscher:2009} approximated a training class with a single Gaussian, and computed the Mahalanobis
distance of a target to the center of the class as an outlier indicator. For the multi-stage approach with multi-dimensional Gaussians,
we use the following extension of the Mahalanobis distance:
\begin{equation}
 d = (a-\overline{\mu})'\overline{\Sigma}^{-1}(a-\overline{\mu}),
\label{distance}
\end{equation}
\noindent with $a$ the attribute vector of the target, and $\overline{\mu}$ the center of mass of the Gaussian mixture. The total
variance $\overline{\Sigma}$ is defined as the sum of the intra-component variances and the inter-component variance:
\begin{equation} 
 \overline{\Sigma} \equiv \frac{1}{N_c} \sum_{k=1}^{N_c} \Sigma_k + \frac{1}{N_c} \sum_{k=1}^{N_c}(\mu_k-\overline{\mu})(\mu_k-\overline{\mu})',
\end{equation}
where $\mu_k$ is the mean vector of each of the $N_c$ Gaussian components. If, and only if, the distance is above a certain threshold, 
the outlier flag will be set to indicate that the target does not seem to belong to any of the predefined classes. This distance is
a multi-dimensional generalisation of the one-dimensional statistical distance (e.g. distance to a mean value of a Gaussian in terms
of the standard deviation). For this reason, a value of the distance threshold d=3 is chosen.

\subsection{Training the classifier}\label{sec:tc}
\begin{figure*}
\includegraphics[width=16cm]{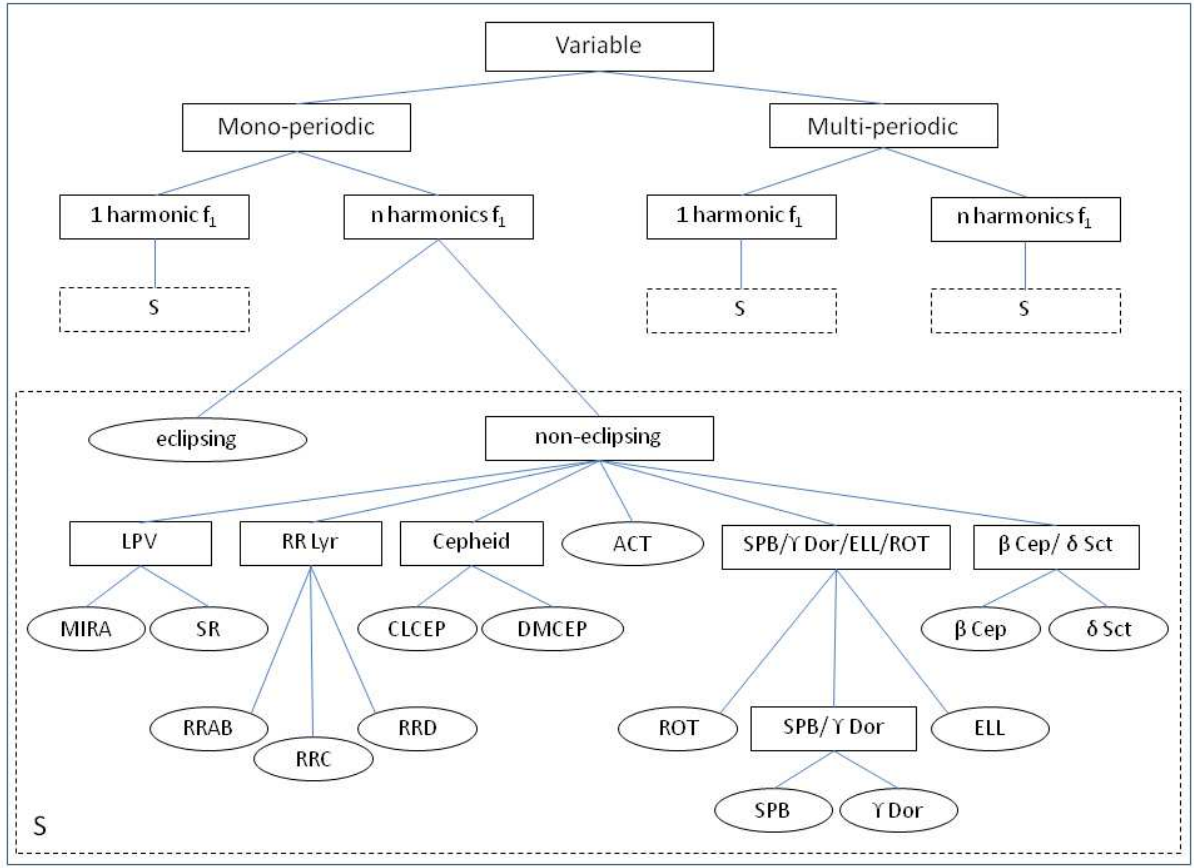}
 \caption{Multi-stage decomposition. The subtree represented in the S box is not replicated for simplicity.}
 \label{mstree}
\end{figure*}

In order to train the classifier, we computed the attributes of the training set objects, which were taken from Hipparcos, OGLE and
CoRoT, with the variability detection pipeline, described in section \ref{sec:vd}. We only computed up to 2 significant frequencies
with each up to 3 harmonics, which in our experience is sufficient for classification purposes. Since the quality of the classification
results depends crucially on the quality of the training set, we checked all the light curves and phase plots in this set. The
variability classes we took into account are listed in Table \ref{classes}. We carefully set up the multi-stage tree, which is given
in Fig. \ref{mstree}. Applying clustering techniques on CoRoT data, \citet{Sarro:2009} managed to identify new classes. In view of the 
\textit{Kepler} mission, two of these classes, stars with activity and variables due to rotational modulation, are taken into account 
in the multi-stage tree. A detailed description of these two classes can be found in \citet{Debosscher:2010}.

\begin{table}
 \centering
 \begin{minipage}{70mm}
  \caption{The variability classes taken into account in the multi-stage tree, with the number of light curves (NLC) used to define 
the classes.}
  \begin{tabular}{@{}lr@{}}
  \hline
   Class     &   NLC\\
 \hline
 Eclipsing binaries (ECL)                       & 790 \\
 Ellipsoidal (ELL)                              &  35 \\
 Classical cepheids (CLCEP)                     & 170 \\
 Double-mode cepheids (DMCEP)                   &  79 \\
 RR-Lyr stars, subtype ab (RRAB)                &  70 \\
 RR-Lyr stars, subtype c (RRC)			&  21 \\
 RR-Lyr stars, subtype d (RRD)			&  52 \\
 $\beta$ Cep stars (BCEP) 			&  28 \\
 $\delta$ Sct stars (DSCUT) 			&  86 \\
 Slowly pulsating B stars (SPB)			&  91 \\
 $\gamma$ Dor stars (GDOR)  		        &  33 \\
 Mira variables (MIRA)                          & 136 \\
 Semi-regular (SR)                              & 103 \\
 Activity (ACT)                                 &  51 \\
 Rotational Modulation (ROT)                    &  26 \\
\hline
\label{classes}
\end{tabular}
\end{minipage}
\end{table}

In each node, we manually selected the best attributes to distinguish the classes considered in that node. In order to evaluate
the significance of an attribute we measured the information gain and gain ratio with respect to each class \citep{WF:2005}. Based on
these results we selected the best attributes in terms of highest information gain and gain ratio, that make sense from an astrophysical
point of view. In practice `random' attributes can show structure, even if they are not supposed to. The attributes we know by theory
that should be random variables were excluded in order to avoid overfitting. In each node the classifier was then tested using stratified
10-fold cross-validation (see e.g. \citet{WF:2005}). In stratified $n$-fold cross-validation, the original sample is randomly partitioned
into $n$ subsamples. Of the $n$ subsamples, a single one is retained as the validation data for testing the model, and the remaining $n-1$
subsamples are used as training data. The cross-validation process is then repeated $n$ times (the folds), with each of the $n$ subsamples
used exactly once as the validation data. Then the $n$ results from the folds are combined to produce a single estimation. Each fold
contains roughly the same proportions of the class labels. We kept the attributes giving the best classification results, not
only in terms of correctly classified targets, but also in terms of accuracy measured by the area under the ROC curve \citep{WF:2005}.
The higher the area under the ROC curve, the better the test.

Stratified 10-fold cross-validation was also applied on the multi-stage tree as a whole. When only the first frequency and its main
amplitude are available, poor results are obtained, because there is simply too little information available for classification. When 
we leave out those examples and only use the training examples for which we have more information, very good results are obtained as can
be seen in Table \ref{cm_tree_gm}. Only $5.8\%$ of the training examples is wrongly classified. When we replace the variability classes
models by single Gaussians, we have a worse result with $7.3\%$ of wrong predictions (see Table \ref{cm_tree_sg}). When we then also use 
only one stage, $10.7\%$ of the training examples is misclassified (see Table \ref{cm_sg}). We can thus conclude that our multi-stage
classification tree with Gaussian mixtures at its nodes, is a significant improvement.

\begin{table*}
 \begin{minipage}{160mm}
 \caption{The confusion matrix for the multi-stage tree applied on the training set objects with at least 2 harmonics for the first
frequency. Each stellar variability class in each node is modelled by a finite sum of multivariate Gaussians. The last line lists the
correct classification (CC) for every class separately. The average correct classification is $94.2\%$.}
 \centering
 \tiny
 \label{cm_tree_gm}
\begin{tabular}{@{}lrrrrrrrrrrrrrrr@{}}
  \hline
      & BCEP & DSCUT & CLCEP & DMCEP & MIRA & SR   & RRAB & RRC  & RRD  & SPB  & GDOR & ELL  & ROT  & ACT  & ECL \\
  \hline
BCEP  & 5    & 3     & 0     & 0     & 0    & 0    & 0    & 0    & 0    & 0    & 0    & 0    & 0    & 0    & 0   \\
DSCUT & 2    & 19    & 0     & 0     & 0    & 0    & 0    & 1    & 0    & 0    & 0    & 0    & 0    & 0    & 0   \\
CLCEP & 0    & 0     & 147   & 0     & 0    & 1    & 0    & 0    & 0    & 0    & 0    & 0    & 0    & 0    & 0   \\
DMCEP & 0    & 0     & 1     & 71    & 0    & 0    & 0    & 0    & 0    & 0    & 0    & 0    & 0    & 0    & 1   \\
MIRA  & 0    & 0     & 0     & 0     & 109  & 5    & 0    & 0    & 0    & 0    & 0    & 0    & 0    & 0    & 0   \\
SR    & 0    & 0     & 0     & 0     & 8    & 61   & 0    & 0    & 0    & 0    & 0    & 2    & 1    & 0    & 1   \\
RRAB  & 0    & 0     & 0     & 0     & 0    & 0    & 67   & 0    & 0    & 0    & 0    & 0    & 0    & 0    & 0   \\
RRC   & 0    & 0     & 0     & 0     & 0    & 0    & 2    & 13   & 0    & 0    & 0    & 0    & 0    & 0    & 0   \\
RRD   & 0    & 0     & 0     & 0     & 0    & 0    & 0    & 0    & 34   & 0    & 0    & 0    & 0    & 0    & 0   \\
SPB   & 0    & 0     & 0     & 0     & 0    & 0    & 0    & 0    & 0    & 19   & 3    & 1    & 0    & 0    & 0   \\
GDOR  & 0    & 0     & 0     & 3     & 0    & 0    & 0    & 0    & 0    & 6    & 6    & 0    & 0    & 0    & 1   \\
ELL   & 0    & 0     & 0     & 1     & 0    & 1    & 0    & 0    & 1    & 0    & 1    & 13   & 0    & 0    & 9   \\
ROT   & 0    & 0     & 0     & 0     & 0    & 0    & 0    & 0    & 0    & 0    & 0    & 0    & 22   & 0    & 1   \\
ACT   & 0    & 0     & 0     & 0     & 0    & 0    & 0    & 0    & 0    & 0    & 0    & 0    & 0    & 46   & 0   \\
ECL   & 0    & 1     & 3     & 0     & 0    & 5    & 0    & 3    & 1    & 6    & 0    & 4    & 3    & 1    & 690 \\
\hline
CC    & 71.4 & 82.6  & 97.4  & 94.7  & 93.2 & 83.6 & 97.1 & 76.5 & 94.4 & 61.3 & 60.0 & 65.0 & 84.6 & 97.9 & 98.3 \\
\hline
\end{tabular}
\end{minipage}
\end{table*}

\begin{table*}
 \begin{minipage}{160mm}
 \caption{The confusion matrix for the multi-stage tree applied on the training set objects with at least 2 harmonics for the first
frequency. Each stellar variability class in each node is modelled by a single Gaussian. The last line lists the correct classification
(CC) for every class separately. The average correct classification is $92.7\%$.}
 \centering
 \tiny
 \label{cm_tree_sg}
\begin{tabular}{@{}lrrrrrrrrrrrrrrr@{}}
  \hline
      & BCEP & DSCUT & CLCEP & DMCEP & MIRA & SR   & RRAB & RRC  & RRD  & SPB  & GDOR & ELL  & ROT  & ACT  & ECL \\
  \hline
BCEP  & 5    & 3     & 0     & 0     & 0    & 0    & 0    & 0    & 0    & 0    & 0    & 0    & 0    & 0    & 0   \\
DSCUT & 1    & 18    & 0     & 0     & 0    & 0    & 0    & 0    & 0    & 0    & 0    & 0    & 0    & 0    & 0   \\
CLCEP & 0    & 0     & 148   & 0     & 0    & 1    & 0    & 0    & 0    & 0    & 0    & 0    & 0    & 0    & 1   \\
DMCEP & 0    & 0     & 0     & 71    & 0    & 0    & 0    & 0    & 0    & 0    & 0    & 1    & 0    & 0    & 2   \\
MIRA  & 0    & 0     & 0     & 0     & 114  & 10   & 0    & 0    & 0    & 0    & 0    & 0    & 0    & 0    & 0   \\
SR    & 0    & 0     & 0     & 0     & 3    & 55   & 0    & 0    & 0    & 0    & 0    & 0    & 0    & 0    & 14  \\
RRAB  & 0    & 0     & 0     & 2     & 0    & 0    & 67   & 0    & 0    & 0    & 0    & 0    & 0    & 0    & 0   \\
RRC   & 1    & 0     & 0     & 0     & 0    & 0    & 2    & 14   & 0    & 0    & 0    & 0    & 0    & 0    & 1   \\
RRD   & 0    & 0     & 0     & 0     & 0    & 0    & 0    & 0    & 32   & 0    & 0    & 0    & 0    & 0    & 0   \\
SPB   & 0    & 0     & 0     & 0     & 0    & 0    & 0    & 0    & 0    & 19   & 2    & 1    & 0    & 0    & 0   \\
GDOR  & 0    & 0     & 0     & 2     & 0    & 0    & 0    & 0    & 0    & 7    & 8    & 0    & 0    & 0    & 2   \\
ELL   & 0    & 0     & 1     & 0     & 0    & 1    & 0    & 0    & 0    & 1    & 0    & 15   & 0    & 0    & 15  \\
ROT   & 0    & 0     & 0     & 0     & 0    & 2    & 0    & 0    & 0    & 1    & 0    & 0    & 23   & 0    & 1   \\
ACT   & 0    & 0     & 0     & 0     & 0    & 0    & 0    & 0    & 0    & 0    & 0    & 0    & 0    & 47   & 0   \\
ECL   & 0    & 2     & 2     & 0     & 0    & 4    & 0    & 3    & 4    & 3    & 0    & 3    & 3    & 1    & 666 \\
\hline
CC    & 71.4 & 78.3  & 98.0  & 94.7  & 97.4 & 75.3 & 97.1 & 82.4 & 88.9 & 61.3 & 80.0 & 75.0 & 88.5 & 100.0 & 94.8 \\
\hline
\end{tabular}
\end{minipage}
\end{table*}

\begin{table*}
 \begin{minipage}{160mm}
 \caption{The confusion matrix for a single-stage classifier applied on the training set objects with at least 2 harmonics for the first
frequency. Each stellar variability class is modelled by a single Gaussian. The last line lists the correct classification (CC) for 
every class separately. The average correct classification is $89.3\%$.}
 \centering
 \tiny
 \label{cm_sg}
\begin{tabular}{@{}lrrrrrrrrrrrrrrr@{}}
  \hline
      & BCEP & DSCUT & CLCEP & DMCEP & MIRA & SR   & RRAB & RRC  & RRD  & SPB  & GDOR & ELL  & ROT  & ACT  & ECL \\
  \hline
BCEP  & 3    & 4     & 0     & 0     & 0    & 0    & 0    & 1    & 0    & 0    & 0    & 0    & 0    & 0    & 2   \\
DSCUT & 3    & 19    & 0     & 0     & 0    & 0    & 0    & 0    & 0    & 0    & 0    & 0    & 0    & 0    & 3   \\
CLCEP & 0    & 0     & 149   & 1     & 0    & 1    & 0    & 0    & 0    & 0    & 0    & 0    & 0    & 0    & 3   \\
DMCEP & 0    & 0     & 0     & 70    & 0    & 0    & 0    & 0    & 1    & 0    & 0    & 1    & 0    & 0    & 4   \\
MIRA  & 0    & 0     & 0     & 0     & 114  & 11   & 0    & 0    & 0    & 0    & 0    & 0    & 0    & 0    & 0   \\
SR    & 0    & 0     & 0     & 0     & 3    & 59   & 0    & 0    & 0    & 0    & 0    & 1    & 0    & 1    & 12  \\
RRAB  & 0    & 0     & 0     & 0     & 0    & 0    & 67   & 0    & 0    & 0    & 0    & 0    & 0    & 0    & 0   \\
RRC   & 0    & 0     & 0     & 0     & 0    & 0    & 2    & 15   & 0    & 0    & 0    & 0    & 0    & 0    & 18  \\
RRD   & 0    & 0     & 0     & 0     & 0    & 0    & 0    & 1    & 35   & 0    & 0    & 0    & 0    & 0    & 4   \\
SPB   & 0    & 0     & 0     & 0     & 0    & 0    & 0    & 0    & 0    & 25   & 2    & 1    & 0    & 0    & 5   \\
GDOR  & 0    & 0     & 0     & 3     & 0    & 0    & 0    & 0    & 0    & 4    & 6    & 1    & 0    & 0    & 8   \\
ELL   & 0    & 0     & 2     & 1     & 0    & 1    & 0    & 0    & 0    & 2    & 2    & 13   & 1    & 0    & 29  \\
ROT   & 0    & 0     & 0     & 0     & 0    & 0    & 0    & 0    & 0    & 0    & 0    & 1    & 23   & 0    & 4   \\
ACT   & 1    & 0     & 0     & 0     & 0    & 0    & 0    & 0    & 0    & 0    & 0    & 0    & 2    & 46   & 0   \\
ECL   & 0    & 0     & 0     & 0     & 0    & 1    & 0    & 0    & 0    & 0    & 0    & 2    & 0    & 0    & 610 \\
\hline
CC    & 42.9 & 82.6  & 98.7  & 93.3  & 97.4 & 80.8 & 97.1 & 88.2 & 97.2 & 80.6 & 60.0 & 65.0 & 88.5 & 97.8 & 86.9 \\
\hline
\end{tabular}
\end{minipage}
\end{table*}

\section{Application to TrES data}
\subsection{The TrES Lyr1 dataset}
We analyzed $25\,947$ light curves in the TrES Lyr1 field. TrES, the Trans-atlantic Exoplanet Survey, is a network of three
ten-centimeter optical telescopes searching the sky for transiting planets \citep{Alonso:2007, Donovan:2008}. This network consisted of 
Sleuth (Palomar Observatory, Southern California), the PSST (Lowell Observatory, Northern Arizona) and STARE (Observatorio del Teide,
Canary Islands, Spain), as TrES now excludes Sleuth and STARE, but includes WATTS. The TrES Lyr1 field is a $5.7^{\circ} \times
5.7^{\circ}$ field, centered on the star 16 Lyr and is part of the \textit{Kepler} field \citep{Alonso:2007}. Most light curves have 
about $15\,000$ observations spread with a total time span of aproximately 75 days. A small fraction has less than $5\,000$ observations
with a total time span of around 62 days. Observations are given in either the Sloan r (Sleuth) or the Kron-Cousins R magnitude (PSST)
and the mean R magnitude ranges from $9.2$ to $16.3$.

\subsection{Classification of variable stars}
\subsubsection{Results of the variability detection}
\begin{figure*}
\includegraphics[width=16cm]{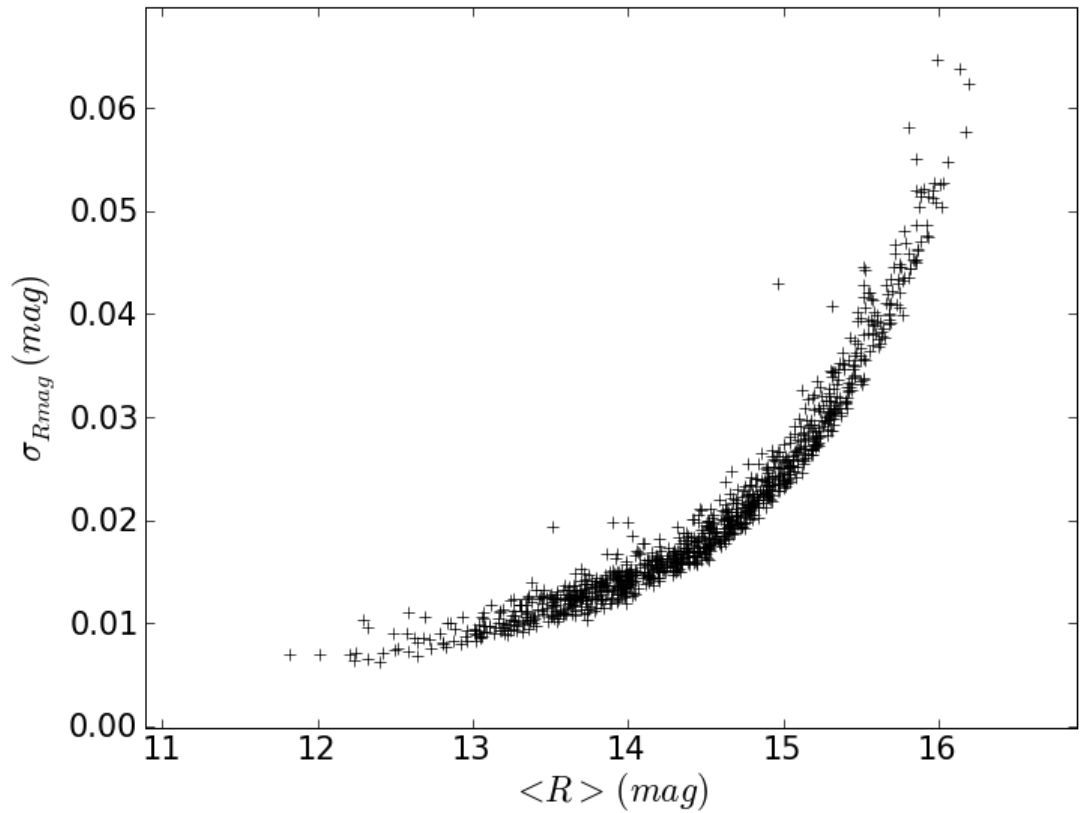}
 \caption{The RMS of the time series plotted as a function of the mean magnitude for stars having no significant frequencies and no
trend.}
 \label{noise}
\end{figure*}

With the use of the variability detection algorithm, described in section \ref{sec:vd}, we searched for frequencies in the range 
$3/T_{tot}$ to 50 c/d, with $T_{tot}$ the total timespan of the observations in days. In order to avoid the problem of daily aliasing
in an automated way, small frequency intervals around multiples of 1 c/d were flagged as ``unreliable''. Using a false alarm 
probability of $\alpha = 0.005$ (the null-hypothesis of only having noise in the light curves is rejected when $P<\alpha$, with $P$
the probability of finding such a peak in the power spectrum of a time series that only contains noise.), about 18\,000 objects were
found non-constant. The stars for which we could not find significant frequencies were used to determine the RMS level of the time
series as a function of the mean magnitude, which is plotted in Fig. \ref{noise}, indicating to what level we can detect variability. 
The upward trend can be explained in terms of photon noise.

\subsubsection{Classification results}
We used the multi-stage tree presented in section \ref{sec:tc}, where we excluded the stars with activity and variables with 
rotational modulation. As already mentioned earlier, these classes were included in the multi-stage tree in view of the 
\textit{Kepler} mission. However, we do not expect to find good candidates in the ground-based data of TrES Lyr1 as these classes are
characterized by low amplitudes. The classification algorithm was able to detect many good candidate class members. By candidate we 
mean a target belonging to the class with the highest class probability above 2 different cutoff values $p_{min}$: 0.5 and 0.75
and with a generalized Mahalanobis distance $d<3$ to that class. A quick visual check of the light curves and phase plots of the
targets with a distance above 3 showed that a large fraction of light curves suffers from instrumental effects. The results of the
classification are listed in Table \ref{candidates}.

\begin{table*}
 \centering
 \begin{minipage}{92mm}
  \caption{Overview of the classification results using 2 different cutoff values for the highest class probability $p$. A generalized
Mahalanobis distance $d<3$ to the most probable class is taken as defined in Eq. \eqref{distance}.}
  \begin{tabular}{@{}lrrr@{}}
  \hline
   Class(es)     &   $p>0.5$  &  $p>0.75$  \\
 \hline
 Eclipsing binaries (ECL)                       & 158 & 130 \\
 Ellipsoidal (ELL)                              & 571 & 214 \\
 Classical cepheids (CLCEP)                     &   3 &   2 \\
 Double-mode cepheids (DMCEP)                   &   0 &   0 \\
 RR-Lyr stars, subtype ab (RRAB)                &   2 &   2 \\
 RR-Lyr stars, subtype c (RRC)                  &   4 &   4 \\
 RR-Lyr stars, subtype d (RRD)                  &   0 &   0 \\
 $\beta$ Cep or $\delta$ Sct stars (BCEP/DSCUT) & 842 & 780 \\
 SPB or $\gamma$ Dor stars (SPB/GDOR)           & 914 & 496 \\
 Mira variables (MIRA)                          &   0 &   0 \\
 Semi-regular (SR)                              &   8 &   5 \\
\hline
\label{candidates}
\end{tabular}
\end{minipage}
\end{table*}

As with CoRoT, the main objective for TrES was the search for planets. We do not find many Long Period Variables (LPV), Cepheids and
RR Lyr among its targets. The total time span of the light curves is also too short to be able to detect Mira type variables.

\subsubsection{Eclipsing binaries and ellipsoidal variables}
\begin{figure*}
\includegraphics[width=16cm]{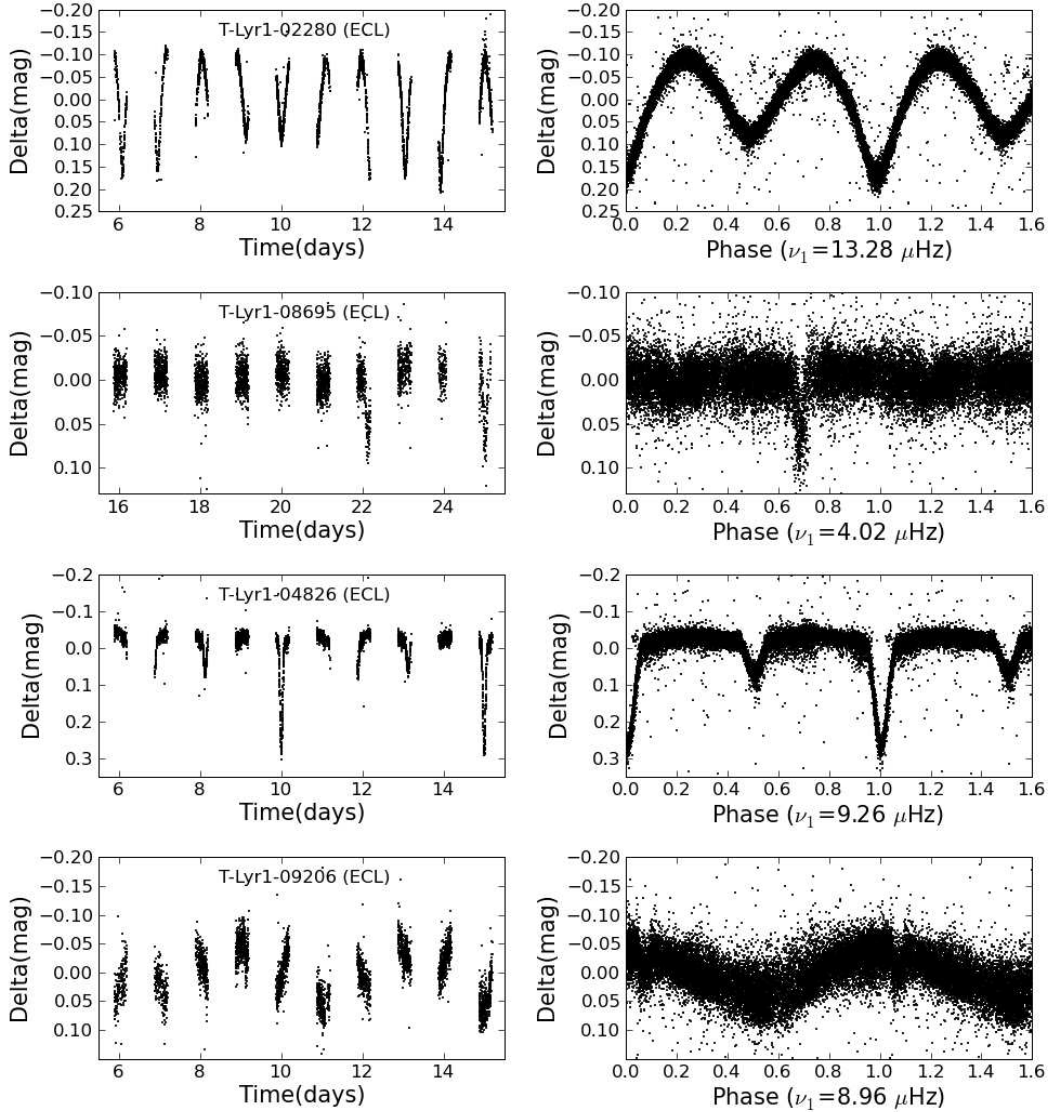}
 \caption{Panels on the left-hand side show a sample of TrES Lyr1 time series of eclipsing binaries. On the right: the corresponding 
phase plots, made with the detected frequency.}
 \label{lc_ecl}
\end{figure*}

Irrespective of the observed field on the sky, we should always find a number of eclipsing binaries and ellipsoidal variables. Light 
curves of eclipsing binaries are very different from those of pulsating stars and therefore generally well separated using the phase 
differences between the first 3 harmonics of the first frequency. Most detected candidate binaries have therefore a very high probability
($>90\%$) of belonging to the ECL class.  We found about 158 reliable eclipsing binaries. Some good examples of eclipsing binary light 
curves are shown in Fig. \ref{lc_ecl}. It is remarkable that, although eclipses are not always easily seen in the light curve, they
clearly show up in the phase plot and are detected by the classification algorithm.

\subsubsection{Monoperiodic pulsators}
\begin{figure*}
\includegraphics[width=16cm]{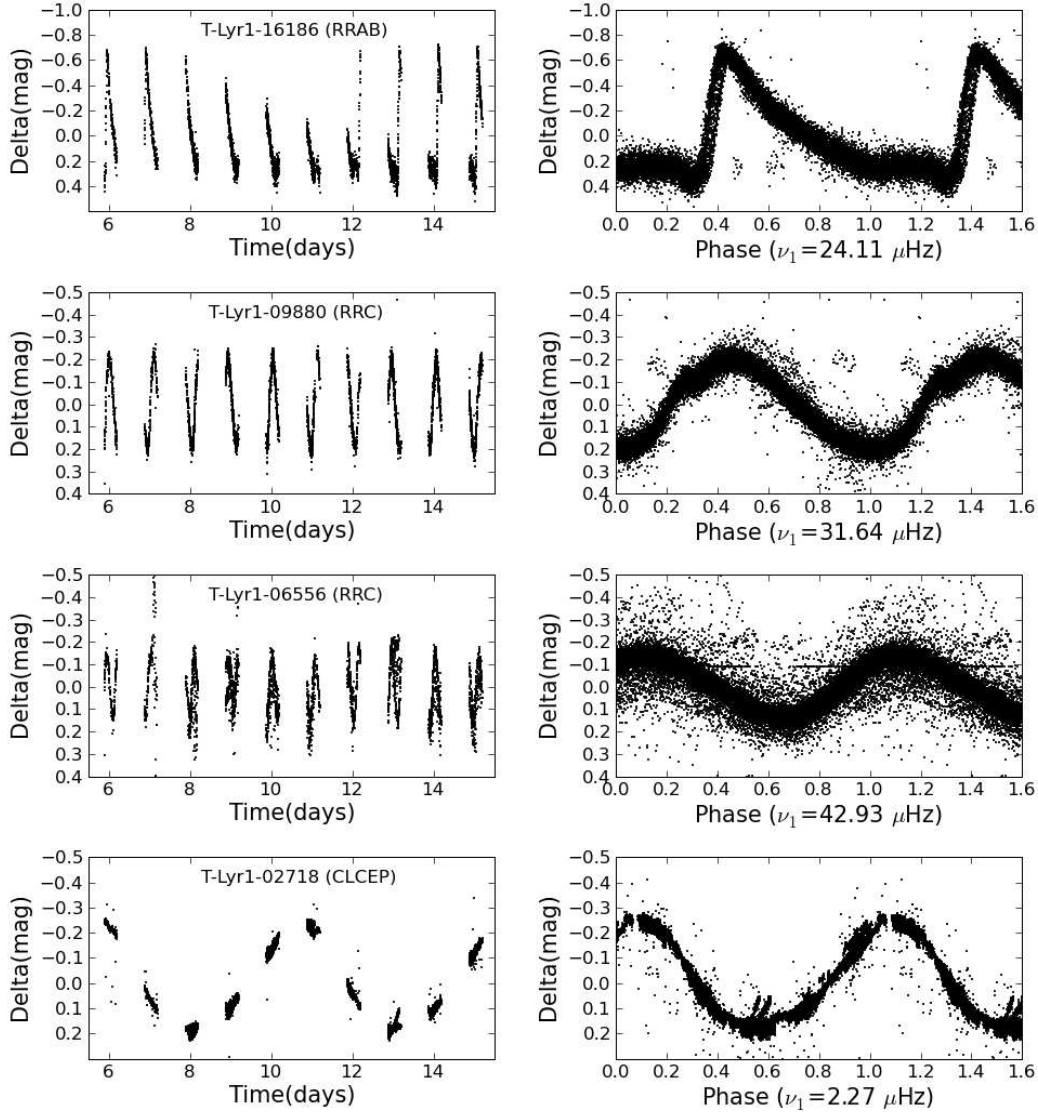}
 \caption{Panels on the left-hand side show a sample of TrES Lyr1 time series of radial pulsators. On the right: the corresponding
phase plots, made with the detected frequency.}
 \label{lc_rp}
\end{figure*}

Despite the fact that Cepheids and RR Lyr are easy to distinguish from other classes due to their large amplitudes, almost no good 
candidates were found. Examples of the few candidates found, are shown in Fig. \ref{lc_rp}.

\subsubsection{Multiperiodic pulsators}
\begin{figure*}
\includegraphics[width=16cm]{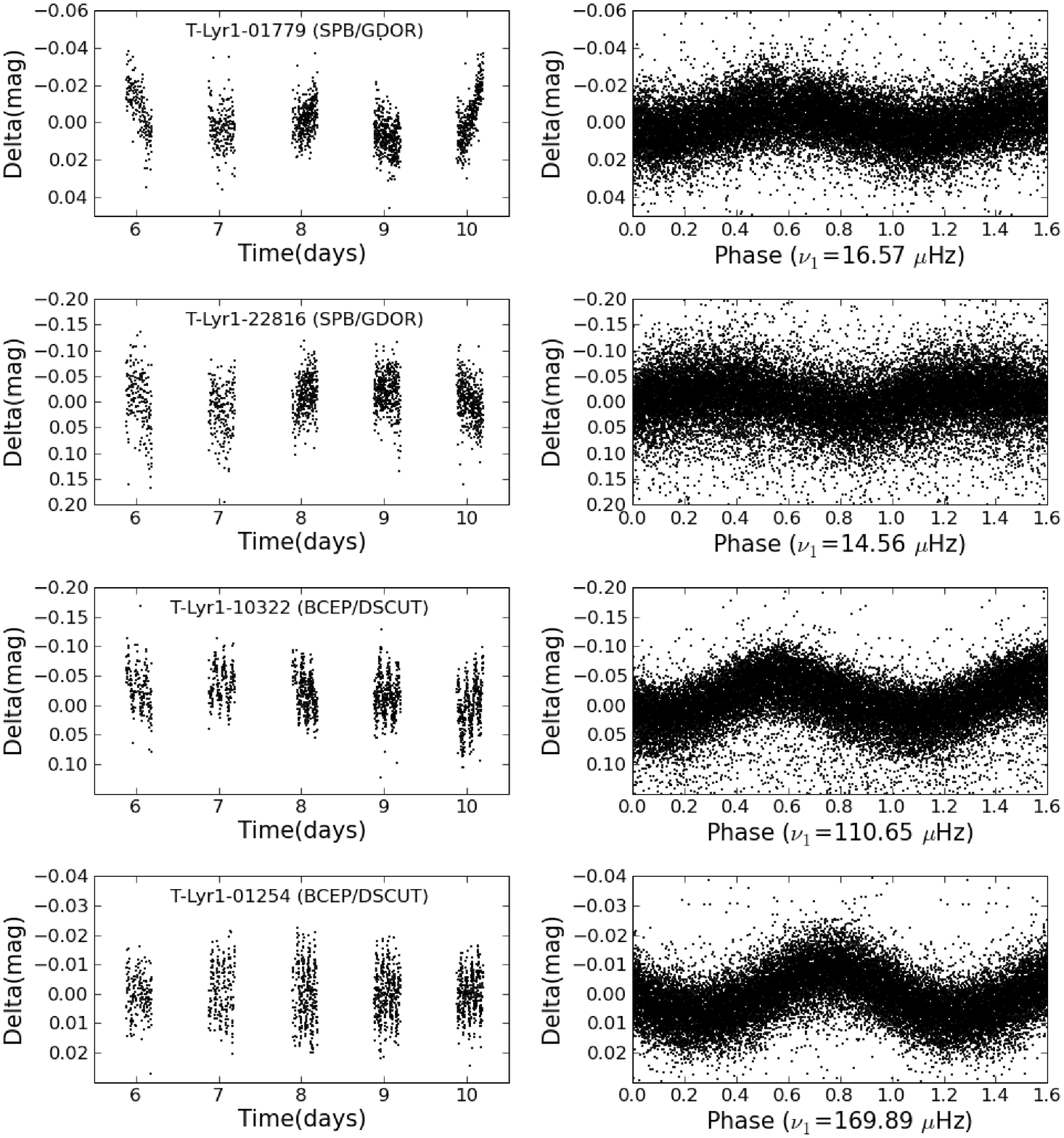}
 \caption{On the left: some TrES Lyr1 light curves of non-radial pulsators. On the right: the corresponding phase plots, made with the
detected frequency.}
 \label{lc_nrp}
\end{figure*}

As no colour information was available, confusion between $\beta$ Cep and $\delta$ Sct stars occurs, because of overlapping frequency
ranges. For this reason we merged these 2 classes into a single class. It is possible that, for the same target, these classes have 
similar probabilities below 0.5, but add up to a value well above 0.5. Similarly, we could often not make a clear distinction between
$\gamma$ Dor and SPB stars, because they show similar gravity-mode spectra. This problem may be solved by adding supplementary
information like temperature, colours or a spectrum, not only for the targets but also for the training sets. Although frequencies
around multiples of 1\,c/d have been set unreliable, especially the $\gamma$ Dor and SPB classes suffer from the combination of daily
aliasing and instrumental effects. For this class, a visual inspection of the light curves and phase plots was needed. Fig. \ref{lc_nrp}
shows some good examples of non-radial pulsators.

\subsection{Discussion and conclusions}
In contrast to previous classification methods for time series of photometric data (e.g. \citet{Debosscher:2007}) we now only use
significant frequencies and overtones as attributes, giving less rise to confusion. We are able to statistically deal with a variable 
number of attributes using the multi-stage approach developed here. Another advantage of this approach is that the conditional
probabilities in each node can be simplified by dropping one or more attributes that are not relevant for a particular node. Moreover, 
in each node, a different classifier can be chosen. In this paper we only used the Gaussian Mixture classifier, but also other methods
like, e.g., Bayesian Nets can be used, which gives more flexibility. Finally, the variability classes were better described by a finite
sum of multivariate Gaussians.

We applied our methods to the ground-based data of the TrES Lyr1 field, which is also observed by the \textit{Kepler} satellite. We found
non-radial pulsators such as $\beta$ Cep stars, $\delta$ Sct stars, SPB stars, and $\gamma$ Dor stars. Because of lack of precise and
dereddened information, and because of overlap in frequency range we could, however, sometimes not avoid confusion between $\beta$ Cep 
and $\delta$ Sct stars, on one hand, and between SPB and $\gamma$ Dor stars on the other hand. Besides non-radial pulsators we also
mention the detection of binary stars and some classical radial pulsators. The results of this classification will be made available
through electronic tables.

\section*{Acknowledgments}
The research leading to these results has received funding from the 
European Research Council under the European Community's Seventh 
Framework Programme (FP7/2007--2013)/ERC grant agreement n$^\circ$227224 
(PROSPERITY), from the Research Council of K.U.Leuven (GOA/2008/04), from 
the Fund for Scientific Research of Flanders (G.0332.06), from the 
Belgian federal science policy office (C90309: CoRoT Data Exploitation, 
C90291 Gaia-DPAC), and from the Spanish Ministerio de Educaci\'on y Ciencia 
through grant AYA2005-04286. Public access to the TrES data were provided to
the through the NASA Star and Exoplanet Database (NStED,
http://nsted.ipac.caltech.edu).


\begin{thebibliography}{}
\bibitem[\protect\citeauthoryear{Aerts et al.}{2010}]{Aerts:2010} Aerts C., Christensen-Dalsgaard, J., \& Kurtz, D.W. 2010, Asteroseismology, Springer-Verlag (ISBN 978-1-4020-5178-4))
\bibitem[\protect\citeauthoryear{Alonso et al.}{2007}]{Alonso:2007} Alonso R., et al., ASP Conf Series, vol. 366, 13]
\bibitem[\protect\citeauthoryear{Debosscher et al.}{2007}]{Debosscher:2007} Debosscher J., Sarro L.M., Aerts C., et al., 2007, A\&A, 475, 1159
\bibitem[\protect\citeauthoryear{Debosscher et al.}{2009}]{Debosscher:2009} Debosscher J., et al., 2009, A\&A, 506, 519
\bibitem[\protect\citeauthoryear{Debosscher et al.}{2010}]{Debosscher:2010} Debosscher J., Blomme J., Aerts C., \& De Ridder J., 2010 A\&A, submitted
\bibitem[\protect\citeauthoryear{Fridlund et al.}{2006}]{Fridlund:2006} Fridlund M., Baglin A., Lochard J. \& Conroy L., 2006 ESA Special Publication, 1306
\bibitem[\protect\citeauthoryear{Gamerman \& Migon}{1993}]{GM:1993} Gamerman D., \& Migon H., 1993. Dynamic hierarchical models. Journal of the Royal Statistical Society, Series B, 55, 629
\bibitem[\protect\citeauthoryear{Gilliland et al.}{2010}]{Gilliland:2010} Gilliland R.L., et al., 2010, PASP, 122, 131
\bibitem[\protect\citeauthoryear{Horne \& Baliunas}{1986}]{HB:1986} Horne J.H., Baliunas S.L., 1986, ApJ, 302, 763
\bibitem[\protect\citeauthoryear{O'Donovan}{2008}]{Donovan:2008} O'Donovan F., 2008, PhD thesis, California Institute of Technology, Pasadena, California
\bibitem[\protect\citeauthoryear{Sarro et al.}{2009}]{Sarro:2009} Sarro L.M., Debosscher J., L\'opez M., \& Aerts C., 2009, A\&A, 494, 739
\bibitem[\protect\citeauthoryear{Schwarzenberg-Czerny}{1998}]{SC:1998} Schwarzenberg-Czerny A., 1998, MNRAS, 301, 831
\bibitem[\protect\citeauthoryear{Witten \& Frank}{2005}]{WF:2005} Witten I.H. \& Frank E., Data Mining: Practical Machine Learning Tools and Techniques (Second Edition), Morgan Kaufmann (ISBN 0-12-088407-0)
\end{thebibliography}
\end{document}